\documentstyle[epsf,12pt]{article}
\newdimen\td \newdimen\tda
\newcount\tc \newcount\tca \newcount\tcb
\newbox\figbox \newdimen\fighgt \newdimen\figwd
\td=\baselineskip \global\baselineskip=\td
\td=\parskip \global\parskip=\td
\newdimen\fighmarg	%horizontal margin between figure and text
\newdimen\figvmarg	%minimum vertical margin between figure and text
\newcount\nindentadj	%user forced extra +/- #lines to indent
\newdimen\dropadj	%user forced extra figure drop or raise (+/-)
\newdimen\sideadj	%user forced extra figure shift right or left
\newdimen\fontht \newdimen\fontdp  %font height and depth
\setbox\figbox=\hbox{A} \fontht=\ht\figbox
\setbox\figbox=\hbox{y} \fontdp=\dp\figbox
\newbox\fontstrutbox
\setbox\fontstrutbox=\hbox{\vrule height \fontht depth \fontdp width 0pt}
\def\fontstrut{%
 \relax\ifmmode\copy\fontstrutbox\else\unhcopy\fontstrutbox\fi}

\def\getfig#1{
 \setbox\figbox=\vbox{\fig{#1}}
 \fighgt\ht\figbox\advance\fighgt\dp\figbox
 \figwd\wd\figbox}

\newif\iffigleft

\newdimen\isize			%indentation size
\newdimen\lsize			%line length
\newtoks\aboveArgs		%args for lines above figure
\newtoks\sideArgs		%args for lines with figure at one side
\def\makeAboveArgs#1{\begingroup  %so \tc is local
\global\aboveArgs={} \tc=#1
\loop \ifnum\tc>0
\global\aboveArgs=\expandafter{\the\aboveArgs 0in\hsize}
\advance\tc by-1
\repeat\endgroup}
\def\makeSideArgs#1#2#3{\begingroup
 \global\sideArgs={} \tc=#1
 \loop \ifnum\tc>0
  \global\sideArgs=\expandafter{\the\sideArgs #2 #3}
  \advance\tc by-1
 \repeat\endgroup}

\def\ifempty#1{\def\next{#1}\ifx\next\empty}  %based on TeXbook, p.263

\def\sideFig#1#2#3#4{
 \ifempty{#2} \tca=0
 \else \tca=#2 \fi			%\tca=#lines above
 \ifempty{#4} \tda=0in
 \else \tda=\parskip\multiply\tda#4 \fi	%\tda=parskip size
 \getfig{#1}
 \advance\tda-\fighgt			%\tda=parskip size - fighgt
 \isize=\figwd\advance\isize\fighmarg
 \lsize=\hsize\advance\lsize-\isize
 \makeAboveArgs{\tca}
 \ifcase#3	%0:  lines above and below
  \advance\tda-2\figvmarg \advance\tda-\fontht \advance\tda-\fontdp
 \or		%1:  flush top, lines below
  \advance\tda-\figvmarg     % \advance\tda-\fontht
 \or		%2:  flush bot, lines above
  \advance\tda-\figvmarg \advance\tda-\fontdp
 \or		%3:  no lines above or below
  \advance\tda\fontht
 \fi
 \tcb=0
 \loop \ifdim\tda<0in \advance\tcb 1 \advance\tda\baselineskip
 \repeat
 \ifcase#3	%0:  lines above and below
  \advance\tcb-1
  \divide\tda 2 \advance\tda#2\baselineskip \advance\tda-\baselineskip
  \advance\tda\fontdp \advance\tda\figvmarg
 \or		%1:  flush top, lines below
  \tda=0in \advance\tda-\fontht
 \or		%2:  flush bot, lines above
  \advance\tda#2\baselineskip \advance\tda-\baselineskip
  \advance\tda\fontdp \advance\tda\figvmarg
 \or		%3:  no lines above or below
  \advance\tcb 1
  \divide\tda 2 \advance\tda-\fontht
 \fi
 \advance\tcb\nindentadj
 \iffigleft \makeSideArgs{\tcb}{\isize}{\lsize}
 \else \makeSideArgs{\tcb}{0in}{\lsize} \fi
 \advance\tca\tcb \advance\tca 1  %\tca=total #args for \parshape
 \advance\tda\dropadj
 \noindent\hbox{\fontstrut}% <-- NECESSARY %!!
 \vbox{\vskip\tda		%drop figure
 \iffigleft \hbox{\hglue\sideadj\box\figbox}
 \else \hfill\hbox{\box\figbox\hglue-\sideadj} \fi
 \vskip-\tda \vskip-\fighgt}	%raise text
 \vskip-\baselineskip \vskip-\parskip
 \parshape\tca\the\aboveArgs\the\sideArgs 0in\hsize
 \global\dropadj=0in \global\sideadj=0in \global\nindentadj=0
}
\def\sideSpace#1#2#3#4#5{
 \ifempty{#3} \tca=0
 \else \tca=#3 \fi			%\tca=#lines above
 \ifempty{#5} \tda=0in
 \else \tda=\parskip\multiply\tda#5 \fi	%\tda=parskip size
 \figwd=#1 \fighgt=#2
 \advance\tda-\fighgt			%\tda=parskip size - fighgt
 \isize=\figwd\advance\isize\fighmarg
 \lsize=\hsize\advance\lsize-\isize
 \makeAboveArgs{\tca}
 \ifcase#4	%0:  lines above and below
  \advance\tda-2\figvmarg \advance\tda-\fontht \advance\tda-\fontdp
 \or		%1:  flush top, lines below
  \advance\tda-\figvmarg     % \advance\tda-\fontht
 \or		%2:  flush bot, lines above
  \advance\tda-\figvmarg \advance\tda-\fontdp
 \or		%3:  no lines above or below
  \advance\tda\fontht
 \fi
 \tcb=0
 \loop \ifdim\tda<0in \advance\tcb 1 \advance\tda\baselineskip
 \repeat
 \ifcase#4	%0:  lines above and below
  \advance\tcb-1
  \divide\tda 2 \advance\tda#3\baselineskip \advance\tda-\baselineskip
  \advance\tda\fontdp \advance\tda\figvmarg
 \or		%1:  flush top, lines below
  \tda=0in \advance\tda-\fontht
 \or		%2:  flush bot, lines above
  \advance\tda#3\baselineskip \advance\tda-\baselineskip
  \advance\tda\fontdp \advance\tda\figvmarg
 \or		%3:  no lines above or below
  \advance\tcb 1
  \divide\tda 2 \advance\tda-\fontht
 \fi
 \advance\tcb\nindentadj
 \iffigleft \makeSideArgs{\tcb}{\isize}{\lsize}
 \else \makeSideArgs{\tcb}{0in}{\lsize} \fi
 \advance\tca\tcb \advance\tca 1  %\tca=total #args for \parshape
 \parshape\tca\the\aboveArgs\the\sideArgs 0in\hsize
 \global\nindentadj=0
}

\def\figNameWrap#1#2{ \def\figprefix{#1}\def\figsuffix{#2} }
\def\fig#1{ \epsffile{\figprefix#1\figsuffix} }

\figNameWrap{}{.eps}
\input epsf
\newcommand{\bge}{\begin{equation}}
\newcommand{\ee}{\end{equation}}

\begin{document}

\begin{center}

November,  1994  \hskip .5truein    PUPT- 1508\\

\vskip .5in

{{\Large    \bf  OSCAR  KLEIN AND  GAUGE  THEORY}}
\footnote{This work was supported in part by the National Science Foundation
under
grant PHY90-21984.}
 \vskip .15in

David J. Gross\footnote{\tt gross@puhep1.princeton.edu}\\[.15in]
{\small {\em
 Joseph Henry Laboratories\\
Princeton University \\
     Princeton, New Jersey 08544 \\}}

 \end{center}

\vskip .15in

\begin{abstract}

In this talk, delivered at the  Oscar Klein Centenary Symposium in Stockholm,
I review the 1938 conference held in Warsaw devoted to  \lq \lq New Theories
in Physics". I review all of the talks presented at this meeting and discuss
in detail Klein's paper where
he proposed a unified model of electromagnetism
and the nuclear force that   foreshadowed the later developments of
non-Abelian gauge theories.

\end{abstract}

 \section{\bf Introduction}

Fifty-six years ago, in September 1938,  there was a remarkable meeting in
Warsaw   devoted to \lq\lq New Theories in Physics"\cite{warsaw}.  This was the
last scientific
gathering which brought together many of the pioneers of quantum mechanics
and the leading lights of theoretical physics before World War II brought an
end to science  as they
knew it.    It was organized by the International Union of
Physics and  the Polish Intellectual Cooperation Committee,  an
organization set up by the League of Nations to promote intellectual
cooperation.  The conference was  held in Poland about a year before the war
broke out
and it was already clear that the intellectual cooperation was beginning to
break
down. Thus, for political reasons, there were no Germans, Italians, or Russians
at this meeting.
 Six years ago   a conference was held in Kazimierz
, just outside of Warsaw, to commemorate the fiftieth anniversary of this
meeting. I was asked   to summarize the
conference. As you know  that is  an awful job, and anyway I did not find the
conference
that interesting so I decided instead to summarize the 1938 conference, which I
found quite fascinating.

The highlight of that
conference, at least with the hindsight of history,  was the remarkable paper
by Oscar Klein in which
he proposed a unified model of electromagnetism and the nuclear force
based on Kaluza-Klein ideas.  This paper stands out
in its originality and its brilliance from the other contributions to the
conference and it   foreshadowed the later developments of
non-Abelian gauge theories that are the foundation of our present theory of
particle
physics.  On this occasion of Klein centenary I thought appropriate to
repeat my summary of the 1938 conference.

The Warsaw   meeting  attracted many distinguished scientists.  The list
of participants included N. Bohr, L. Brillouin, L.  de Broglie, C. Darwin, A.
Eddington, R. Fowler, G. Gamow,
S. Goudsmit, Oscar Klein, H. Kramers, L. de Kronig, P. Langevin, C. Moeller,
J. von Neumann, F. Perrin, L. Rosenfeld, and E. Wigner.
 There was a report  from Heisenberg who did not
attend, presumably he was not allowed to go to  Poland, and from E. Milne.
I  noticed that the three participants who came from my home
institution of Princeton (Goudsmit, Von Neumann  and Wigner) were all part of
the large gift of the Nazis to
American science.   The scientists  sat around a round table, which  stimulated
  discussions. One of the nice features of  this
meeting was that there were  only nine talks
but  extensive discussions,  recorded
by industrious graduate students and postdocs. Thus  the proceedings contain
the
 discussions which were at times much more interesting than the
talks.

  I  shall review the  whole conference,  which consisted of
  nine talks,   discussing  each of them in term and   Klein's  contribution in
detail.  I will
also try to draw some lessons from this remarkable
episode.  If I seem to emphasize
the confusion and the errors made by many of the heroes of modern science
at this meeting (including Klein),  an easy task to do
with the hindsight of fifty years,  it is not out of
disrespect for these giants of modern physics but rather for two
reasons:
\begin{itemize}
\item  To put Klein's contribution in perspective.  By discussing
some of the other talks,   sometimes in humorous tones, this will make
Klein's contribution look even more remarkable.

\item    I think it is
very important in studying intellectual history not  to indulge in hero
worship.
History is not just an account of the great triumphs and successes of the
past but also of the false leads and errors and mistakes that our heroes
made.  Only if we learn about these can we truly appreciate their triumphs.
Only by studying the false paths they sometimes followed can we begin to
appreciate them as real human beings and not as gods.
\end{itemize}

I shall    start   by reviewing all of the other  talks and then we will move
on to Oscar Klein.

\section{\bf Review of the Conference}
\subsection{\bf Bohr and von Neuman}

After an opening address by Professor Bialobrzeski
the first talk  was given  by Niels Bohr, followed
by  a contribution from John  von Neuman.  They   discussed
interpretational issues  in quantum mechanics.  The title of Bohr's talk was
\lq \lq The Causality Problem in Atomic Physics''.  Bohr presented  a very
clear and comprehensive review of his views of complementarity and the
Copenhagen
interpretation of quantum mechanics.  In this paper he gave, for the first
time,  a
  precise definition of  what a physical phenomenon meant, namely that
\begin{quote}
{\sl  One should reserve the word phenomenon for the
comprehension of the effects observed under given experimental conditions.}
\end{quote}
Bohr presented   a very clear   exposition of the Copenhagen interpretation of
quantum mechanics but did not discuss any  new theories. von Neuman,
who spoke after Bohr, talked about two of his recent papers.  One was a
proof that there could be no hidden variable explanation of quantum
mechanics. He   presented the proof of this assertion.  von Neuman's hidden
variable theorem was wrong,  as was discovered  many years later  by David Bohm
who constructed a
consistent hidden variable theory and by  John Bell
who pointed out von Neuman's mistake.  The second of von Neuman's
contributions was a
discussion of some work he had done with the mathematician Birkhoff in which
they tried to understand quantum mechanics   by changing the rules
of ordinary logic, replacing the structure of the propositional calculus of
logic based  on Boolean algebra by one based on
the  properties of rays in a
Hilbert's space.  This was not  wrong, but Bohr  objected strenuously. He
remarked that
\begin{quote}
{\sl Personally, he compelled himself to keep the logical
forms of daily life.}
\end{quote}

There followed a discussion,  mostly of quantum
logic.  According to the rules of  meeting only the invited speakers,
  were allowed to  talk during the discussion periods. None of the younger
members
of the audience, in particular the many Poles who were present, were
allowed to open their mouths.  The only exception to this rule was a
Monsieur Destouches from Paris  who was allowed to talk, and in fact   talked
at great length after every talk except for Klein's.
 Destouches was a protege of de Broglie and he occupied a position of some
eminence
in France.  His contributions to physics were  summarized
by  Abragam, who  stated in his Memoirs that, as a student growing up in the
French
scientific environment, he felt it necessary to study  Destouches's papers.
\begin{quote} {\sl
I struggled very long to understand
until I understood that there was nothing to understand.}
\end{quote}
Destouches contributed  a long comment on von Neuman's quantum logic and, since
I did
not struggle to understand, I will not summarize his remarks.

\subsection{\bf  Louis de Broglie}

The second talk was by de Broglie. Since de Broglie   did not attend in person
his  contribution was read by   E.  Bauer and  was supplemented by Destouches.
de Broglie's  talk was entitled
\lq\lq Links Between the Quantum Theory and Relativity''.  He discussed the
difficulties of reconciling quantum mechanics and relativity.  What were these
difficulties?  de Broglie first remarked that Pierls and Landau have showed
that
one cannot define  position to better than the Compton wave length of the
electron.
Spacetime, he said, is an idea drawn from large scale experience.
And here we have a limitation on spacetime.  Furthermore as he noted, the
usual Hamiltonian quantization techniques treat time and space
asymmetrically, which is in conflict with relativity and this bothered him.
These concerns  were then  amplified by Monsieur Destouches, who described at
length
a totally bizarre relativistic particle dynamics of his invention in which
each particle has its own time. He also described    a theory of de Broglie
which, as far as I
can tell, had nothing to do with the previous issue, in which the photon was
to be thought of as a composite of two spin-1/2 massless particles as if   the
photon was a composite   of two  neutrinos.

This paper provoked much criticism.  Bohr  clarified very
clearly and concisely that:  (1) the problem of localization, i.e., of
measuring the position of a particle to better than its Compton wave length,
disappears completely if you admit the reality of negative energy solutions
as was explained, he says, by Klein ; and (2) the problem of asymmetry of space
and time is, as we all know today, merely a technical problem  that
could be dealt with;  but in any case does not conflict with relativity.

\subsection{\bf Werner Heisenberg}
The next talk was that of Oscar Klein, to which I shall return later.
Following Klein was the contribution of   Heisenberg. He too was not present,
so his contribution was   read
by Kramers.  Heisenberg discussed the limits of the applicability of the
present system of theoretical physics.  What were the  problems as he saw them?
 He
identified two major issues concerning theoretical physics at that time.
The first was the existence of ultraviolet divergences in  quantum field
theory. What
he meant at that time by ultraviolet divergences was the self energy
of the electron--nothing more.  The second problem that  concerned him was
the experimental observation of
cosmic ray showers, in particular the occurrence
of particle production  in these high
energy showers.  Heisenberg concluded from the existence both of ultraviolet
divergences and  multi-particle production   that there had to
be a fundamental length of order  the classical radius of the electron, below
which the concept of length loses its
significance and quantum mechanics breaks down. The classical electron radius,
$ e^2/mc^2$  is clearly associated with the divergent electron self-energy, but
also
happens to be the range of nuclear forces, so it has something to do with
the second problem. Quantum mechanics itself, he said,
should break down at these lengths.
I have always been  amazed at how willing the great inventors of quantum
mechanics
were to give it up all at the drop of a divergence or a new experimental
discovery.

 George Gamow was present as well and he gave a short presentation of an
alternate
explanation of cosmic ray showers that did not require giving up
introducing a fundamental length. His explanation was simply that nuclear
forces were described by  Fermi's
theory of beta-decay. He wrote down a formula for the cross section
for the production rate
of particles in Fermi's theory which would go like the energy to the fifth
power.
This  is wrong but we do know that the cross sections in Fermi's theory
of the weak interactions do increase with energy and he realized that.  So
maybe one could explain why the probability of  producing many particles would
increase at high energies and thus explain the multiparticle production in
the cosmic ray showers.  However, he noted that there is a slight problem;
namely in
order to account for the proton-neutron interaction as well as the showers
one  requires that the coupling be of order $1$ instead of Fermi's coupling,
so that one is off by a factor of $10^{12}$.  But he still presented the idea.

\subsection{\bf L. Brillouin}
 The next talk after  was one of the most interesting of all. It
was a talk by Brillouin called \lq \lq The Individuality of Elementary
Particles''.
 I suppose he was asked to talk about statistics.  Instead  he gave
  a long review of   the present state of what we call today
elementary particle physics.
I find this talk,   aside from Klein's, to be the most interesting at the
conference
since it  describes  what people knew about particle physics in 1938
 and what they regarded as the important problems.

The first thing Brillouin
he showed was a table of the elementary particles as known at the time:

\begin{tabular}[t]{ |l@{}   |r@{ }|r@{ } |r@{ } |}
\hline
\multicolumn{4}{|c|}{\bf  Table I. ------  Elementary  Particles} \\
\hline \hline
{ \bf Particle    }
& {\bf Mass at rest    }
& {\bf Charge    }
& {\bf Spin     }\\
\hline \hline
  Electron
& $m_0$ & $-e$ & $1/2$ \\
 \hline
Positron & $m_0$ & $+e$ & $1/2$ \\
\hline
Heavy electron
& $100 m_0$  & &  \\
 \hline
Barytron, Mesotron  {}
& $200 m_0$  & $\pm e$ &$1$  \\
 \hline
 Neutron
& $M_n$  &$0$  & $1/2$ \\
 \hline
 Proton
& $M_p$  &$+e$  & $1/2$ \\
 \hline
 Photon
& $0$  &$0$  & $1\,  ({\rm or}  \, \,0?)$ \\
 \hline
  Neutrino
& $0$  &$0$  & $1/2$ \\
 \hline
\end{tabular}

\bigskip
 The table consisted of  the electron,  as well as the   positron
which had already been discovered, the proton and
the neutron, the photon, the  neutrino (which Brillouin identified with   the
anti-neutrino since it had no charge),      heavy
electrons and   mesotrons.

This list contains some strange entries.  First   there was much  confusion as
to   the nature of the particle that had been recently
observed in cosmic rays.  Everyone assumed  that it was
  the particle that Yukawa had proposed as   mediating
the nuclear force.  However, as we now know there were two new particles in the
cosmic ray events, the pion as well as the muon. This  confusion
is evident in the list and Brillouin 	 refers sometimes to a
  heavy electron and sometimes to 		as a mesotron.  The mesotron,
 Yukawa's particle, he writes has spin one.  Why?  Yukawa
originally supposed, he says, that the spin was equal to zero but later
\lq \lq calculations determined the spin to be $1$".  He does not explain what
those
calculations were.  Presumably they were the fact that Proca had suggested
a wave equation for a spin-1 particle.  It was very unclear at that time
which  equation one should use to describe a given particle.

The most fascinating thing in this list is the treatment of the photon.
Brillouin
says that  \begin{quote} {\sl   The
photon  represents a daring abstraction,  for it does not possess charge or
mass when at rest. } \end{quote} Thus  in  1938, 33 years
after  Einstein's proposal  of the photon, it was still a daring abstraction.
  As to
its spin, it was formally supposed to be nil, he says.  But if it obeys a
linear wave equation, then the spin should be one.  And again, it was
unclear to him, although not to Kramers, who gave a very nice retort in
the discussion period, how you describe the wave equation of the photon.
So this was the list of elementary particles.

What were the outstanding problems of particle physics? The    first problem
was the stability of the electron.
Why is the electron stable ?  One theory that might deal
with this problem,  according to Brillouin, was the nonlinear theory of Born
and Infeld in which the Maxwell
Lagrangian is replaced by  $L= b^2  \sqrt{1+(B^2-E^2)/b^2}$, which reduces to
Maxwell's Lagrangian when  $b \to 0$.  How this theory    solves the stability
of
electron was not explained.

The second problem was which wave equation to use for each particle.    One had
available the   Dirac equation, the Klein-Gordon equation (which he    called
the
Gordon-Maxwell equation), Proca's equation and so on.

Then there   is a long discussion
of super quantization, which nowadays we call second quantization.  This
discussion is a  marvelous illustration of how confused people
were about the   new quantum field theory. It was unclear to Brillouin
whether
second quantization was something which went beyond quantum mechanics of
the usual type or not or  whether it was necessary.  He
states quite clearly that second quantization is only necessary for
particles obeying Bose statistics, with spin zero or spin one. Particles
obeying Fermi statistics with half integer spin,  that  obey linear wave
equations, do not require second quantization but  can be treated by the
hole theory of Dirac.  Clearly there was no understanding that these  two
approaches were
equivalent.

Brillouin also discussed
  the nuclear force.  This discussion, as well as Gamow's
earlier remarks,
illustrates that at the time there was absolutely no understanding that there
were two forces under
discussion, that there was any difference between the interactions that gave
rise
to beta decay and the forces that held the nucleus together and gave rise
to neutron proton scattering.   There
was an enormous amount of confusion  as to whether one should describe the
nuclear force using
  the Fermi interaction or   Yukawa's idea of  a  meson induced force.  Finally
there
was a long discussion of de Broglie's idea that the photon should be thought of
as a neutrino and by neutrino pair.

There was a lot of discussion after this  contribution. Some of the confusion
was, or should have been, dispelled by  Kramers.   One must say that Kramers
was the most intelligent participant in the discussion sessions.

\subsection{\bf Arthur Eddington}

Following Brillouin that there was a talk by Eddington.  This is one of the
most
remarkable episodes in the whole meeting. Eddington  was
a famous English astronomer who had made Einstein famous by observing the
predicted deflection of light by the sun.  He was a great astrophysicist and a
great
popularizer of science.  As I child I remember reading his books.  There were
very well written, wonderful  popular
science.  But at some point he  over reached himself and thought he had a
theory of everything including a precise determination of the fine
structure constant (his theory gave $\alpha = 1/136$---good to $1\%$), the
radius of the universe,   the ratio of all masses, etc.
 Eddington's talk was entitled  \lq \lq The
Cosmological Applications of the Theory of Quanta."  He  took it
for granted that everyone accepted the fact that he could calculate the value
of
$\alpha$. He presented his calculations of the number of
particles in the universe and the radius of the universe and so on.  Thus the
number of particle in the universe is $3.145 10^{79}\approx 2\times 136\times
2^{256}$ and the radius of the universe is $1.234 10^{27}{\rm cm.} $  The
theory is
totally incomprehensible.

After  Eddington's talk there was a  very long discussion session.
Eddington was   the Carl Sagan of his time,    a very popular figure
with the media. He published  his  theories of everything, but not
in scientific journals. He never appeared at scientific meetings and
all of the scientists resented  him for his publicity seeking and lack of
critical scientific attitude.  This
was the first time he had  ever talked about these theories to a
scientific audience.   Many in the audience were  waiting to ambush him.
Everyone jumped on him, including
Kramers, von Neumann, Rosenfeld, Wigner, Gamow, Fowler and Bohr.
Everyone said, very  politely,  that the way he approaches all parts of
physics, including  quantum mechanics and relativity, is in contradiction
with the ordinary theory of quantum mechanics and  relativity.

Kramers was  elected by the younger members, especially Gamow, to deal with
Eddington and he gave the longest discussion in which he criticized Eddington's
views.  When I was   at the anniversary  meeting  in Kazimierz  the
organizers showed me an illustration, that came  from their private files, in
the form  of  a medal that Gamow presented to
Kramers after he had performed this service to the community. The medal
reads:\quad
\lq\lq  For the masterpiece of polite scolding."   For most of the
the participants this talk and the following discussion  was the highlight of
the meeting.

\subsection{\bf A. E.  Milne}

Following Eddington  came the contribution of the cosmologist E. Milne.   Milne
was not present  so his
contribution was read by Darwin---a very respectable physicist.  Milne gave a
talk on \lq \lq A Possible Mode of Approach to  Nuclear Dynamics",  which was
even crazier than
Eddington's.  He introduced some sort of absolute time based on  Mach's
principle,    gave up conservation of energy and
momentum and then  he deduced Coulomb's law and the Bohr orbits and so on. When
Darwin finished presenting the talk of his colleague, he stated that \lq \lq
having read Professor Milne's paper, he  wished to say
  that he did not agree with the conclusions of the paper or
certain of the assumptions in it."  Since Milne was absent, there was
no discussion of the paper.

\subsection{\bf Paul Langevin}

The last talk,  by  P. Langevin,  was entitled  \lq \lq On the Positivistic and
Realistic Trends in the Philosophy
of Physics."  It was  philosophy  and not
physics----positivism versus realism.  It is hard for me to read
this kind of stuff.  As far as I can tell, realism won.

Finally the meeting ended with a comment of the chairman, C. Bialobrzeski.
After thanking the participants he  remarked about the great contributions of
modern science as indicated by the energy theory of Wilhelm
Ostwald who   showed that energy was the primordial substance. He stated that:
\begin{quote}{\sl The chief
advantage of   the energy theory is  that this
doctrine bridges the gulf that separates physical and psychic phenomena.}
\end{quote}
This   gives you some idea of the background to Oscar Klein's
contribution to the meeting.

\section{\bf Oscar Klein's Theory}

Oscar Klein gave the fourth talk entitled \lq \lq On the Theory
of Charged Fields.''  He started by explaining the motivation for the theory.
The  primary motivation  was Yukawa's meson hypothesis, made in 1935 and
recently  confirmed by experiment.  This proposal
of Yukawa and  its rapid  experimental confirmation had an
enormous impact on theoretical physics,  certainly on Klein. Yukawa proposed
that the force between protons and neutrons  was  mediated by a  meson in the
same way that the electromagnetic force is mediated by the
photon,  except that Yukawa's  meson was very  massive. If the meson mass was
of order  100 MEV then one could   explain the short range nature of the
nuclear force.
The evidence in cosmic rays for a particle that might fit this role
 came very shortly after the
proposal was made.

Klein stated  that Yukawa's  idea and its confirmation implied  {\em a
considerable
enlargement of the field concept.}  What did he mean?  The  paradigm of a
quantum field theory at that time was quantum electrodynamics.
The developers of quantum electrodynamics, including Klein, knew
that the theory had severe ultraviolet divergences.  The divergences that they
focused on   were the self energy divergences. They believed that these
divergences meant that the theory  must be altered at distances  smaller than
the
  Compton wave length of
the electron $=\hbar/mc\approx 10^{-11}$ cm. (Remember Heisenberg's paper.)
Mesotron dynamics, according to  Yukawa,  involved
 a particle that  is about $100$ times heavier than the electron.
Therefore, Klein notes, mesotron dynamics can work down to a much smaller
distance  of order the  Compton wave length of  the pion.  Thus if we
incorporate the mesotron field
  we might extend the framework  of quantum field theory by two orders
of magnitude farther, from $10^{-11}$ cm., in the case of   QED by itself, to
$10^{-13}$ cm. in the case of the nuclear force.  Furthermore, Klein noted that
if
we combine electromagnetism with the nuclear force  we might somehow  be able
to understand the self energy problem; in fact we might be able to
understand the rest mass of the electron.  After all  the mass
of the electron might just be  Coulombic in origin,  if the characteristic
distance scale is   set by the heavy meson mass.  Since the ratio of the
electron mass, $M_e$ to the meson  mass $M_m$ is
of order $\alpha$, the Coulomb   potential at a distance of order the   heavy
meson  Compton wave length is   of order  the rest mass of the electron,
$\alpha M_m \approx M_e$.  That is what Klein meant
  by \lq \lq {\em a reasonable enlargement of the field concept. }"

What was Klein's  goal?  His  goal was
very ambitious.  It was nothing less than a theory of everything, but in a
much more realistic sense than Eddington.  He wanted to construct a field
theory that described   all the matter that that was known to exist,
namely the neutron, the
proton, the electron and the neutrino, interacting with the fields he thought
are necessary to give all the forces that were known---electromagnetism  and
the
nuclear force.
Thus he wanted a theory of
\begin{equation}
{\rm Matter: } \ \left( \matrix{n \cr p\cr } \right)  + \left( \matrix{\nu \cr
e \cr } \right)\ , \ {\rm Interacting \ with\ \ } \ \
\left( \matrix{{\rm Electromagnetic\  Field}  & \gamma \cr {\rm Mesotron \
Field} & M^\pm \cr} \right)
\end{equation}
 Like everyone else he did not distinguish  between the weak and the
strong interactions, both were to be described   by the mesotron field  of
Yukawa.  Thus
his goal was  a complete and  unified theory of electromagnetism plus the
nuclear forces,
and since the theory was based on   a gravitational context,   gravity as well.
 This was perhaps the first respectable attempt to construct
  a {\em theory of everything.}

How did Klein go about constructing  this theory?  The method he followed, not
surprisingly given the history of Klein's involvement with the unified theory
of electromagnetism and
gravity, was to use what he called the five dimensional representation.  This
of course was the Kaluza-Klein theory which   explained electromagnetism in
terms of a five dimensional theory of gravity, where the fifth dimension was
compactified on a small circle. The
main advantage of this approach, according to Klein,   was that it
automatically preserved energy-momentum conservation, charge
conservation, and gauge invariance.
Since he wanted to construct a new theory   he decided to use this formalism
which automatically preserved the
symmetries. But the five dimensional theory was already constructed,  so what
was new?  The new ingredient---which explains the title of his talk---was that
he wanted to describe charged gauge mesons and therefore he  included in the
theory, for the first time,
$x^5$-dependent fields. ($x^5$  denotes   the fifth
dimension, a little circle of radius the Planck length $ \approx  10^{-33}$
cm.)   A field  which has a non-trivial dependence on $x^5$  will carry
quantized five-momentum.   The fifth component of the momentum is quantized  in
units of the inverse  radius  and couples to   to the long range five
dimensional gravitational field. At low energies this appears like a charged
particle coupled to the  electromagnetic field.      Thus  by making the fields
$x^5$-dependent one can describe charged particles. Klein needed to describe
charged  particles, both matter fields, such as the proton  and the electron,
and force fields, such as the mesotrons.   In particular the five dimensional
metric tensor field, which to
a low energy observer looks like a four dimensional metric tensor field plus a
vector
meson field and a scalar meson field, will now have $x^5$-dependence. He
ignores the charged graviton and the dilatons
 but identifies  the charged gauge bosons    with
Yukawa's mesotrons.  The Dirac spinors that he introduces to describe the
matter will also contain $x^5$-dependent pieces that  will be used to describe
describe the proton and the electron.
Note that Klein was not  trying to increase the symmetry of the world.  There
is no
discussion  of a new $SU(2)$ symmetry or  of enlarging the notion of gauge
invariance.

Let me reiterate.  Klein's goal was a theory of everything---a five dimensional
theory of gravity plus electromagnetism plus the nuclear force--all the
forces known at the time interacting with  all of
the matter known at the time. The matter he puts into two   families, the
the proton and  neutron multiplet
  and the electron and  neutrino multiplet. This is the first time that
families are introduced.    Klein notes the fact these multiplets are
repetitious, much like the quark-lepton families of the standard model  and he
adjusts the mass (as we do today)
to account for their mass differences.
 The parameters of his unified theory
consist of the    electric charge and the mass of the
proton and neutron.  The mass of the electron and
the neutrino he takes to be zero.   He imagines that electron  mass will come
emerge dynamically.  He also adds a mass term for  the mesotrons.  Such a mass
term violates the gauge invariance of the nonabelian
Yang-Mills theory of these gauge bosons.   But he was not trying to construct
an $SU(2)$ gauge
invariant theory and in fact he did not.  It clearly bothered him to have to
introduce a mesotron mass term by hand and he states that
\begin{quote} \sl
It is not impossible that a further development of the theory will make this
somewhat arbitrary addition superfluous, the mass appearing as some sort of
self energy determined by the other lengths entering in the theory.
\end{quote}
That of
course is the way it works in the real world as we understand it   now--the
masses of the $W$ and the $ Z$ mesons
are not introduced by hand but generated dynamically.
But Klein was not aware that these  explicit mass  terms violated gauge
invariance. To the contrary, he states
\begin{quote}\sl
As to the rest mass of the new particle, which does not appear in the ordinary
field equations, it might be introduced by the addition of a term in the
Lagrangian without disturbing the invariance. \end{quote} The reason was that
he was
not thinking of $SU(2)$ gauge  invariance at all.

I shall now describe the theory that Klein constructed.
  He  starts with the matter sector much as   Yang and Mills did in deriving
Yang-Mills theory.  Yang and Mills started with an isotopic spin
doublet and tried to render  that theory of isotopic spin doublets gauge
invariant,  inventing  the Yang-Mills gauge bosons to do so.  Klein   starts
with
the  neutron and the proton which he puts together in an
isodoublet,  following  Heisenberg, although he never refers to   isotopic spin
symmetry.   He puts the neutrino and the electron in a second isodoublet, which
 he remarks is just a repetition of the first.  Both isodoublets are described
as five-dimensional Dirac spinors.
The proton and the electron acquire their charge  through the  $x^5$-dependence
of the fields.  Since $x^5$   is
canonically conjugate to the fifth component of the  momentum which is
identified with
electric charge, the $x^5$ derivative of $\Psi$  vanishes for the neutron
and yields $e$ for the proton, and the same for the electron-neutrino
multiplet.
Thus the matter fields are:
\begin{equation}
\Psi_1 = \left( \matrix{\psi_n \cr \psi_p\cr}\right), \ \ \Psi_2 =\left(
\matrix{\psi_\nu \cr \psi_e\cr}\right);\
\ {\partial\over \partial x^5} \Psi_1 =  {ie \over \hbar c}\left( \matrix{0 \cr
\psi_p\cr}\right),\ {\partial\over \partial x^5} \Psi_2 = -{ie \over \hbar
c}\left( \matrix{0 \cr \psi_e\cr }\right) ,
\end{equation}
and Klein took the  Lagrangian  to be the relativistically
 covariant   Dirac Lagrangian in five dimensions. (Five dimensional  spinors
were introduced previously by  Schrodinger.)    He   adds a mass term by hand
and
arranges the mass of the proton and the neutron to be identical   and for the
mass of   the electron and  neutrino to be zero.
\begin{equation}
{\cal L} = \bar \Psi i \gamma^{\hat \mu} \partial_{\hat \mu}\Psi + \bar \Psi M
\Psi
=  \bar \Psi( i \gamma^{  \mu} \partial_{  \mu} + \sqrt{\kappa} \chi_\mu
\gamma^{  \mu} {\partial \over \partial x^5})\Psi + \bar \Psi M \Psi + {\rm
gravitational \ \ pieces}
\end{equation}
 Following
the usual Kaluza Klein philosophy,  when   the five dimensional Dirac
Lagrangian is  written in four dimensions one  gets,  in
addition to the  usual Dirac Lagrangian, a   term that  looks
like the  minimal coupling of a gauge field $\chi_\mu$ to the electric current.
 In this
case, since Klein starts with isodouplets, the gauge field is  as a $2\times 2$
matrix with diagonal     components $A_\mu$, proportional to the neutral
components of $g_{\mu 5}$
 and  off diagonal components $B_\mu$, a complex  field  with an  $x^5$
dependence on  corresponding  to the  $\pm  1$
charged  components of  $g_{\mu 5}$,
\begin{equation}
\chi_\mu = \left(  \matrix{ A_\mu & \bar B_\mu \cr B_\mu  & A_\mu \cr
}\right);\ \
{\partial \chi_\mu \over \partial x^5 } ={ie\over \hbar c} \left(  \matrix{ 0 &
-\bar B_\mu \cr B_\mu  &0 \cr }\right) .
\end{equation}

 The $A_\mu$ fields are identified with the electromagnetic field and the
$B_\mu$ fields with    the mesotrons
that mediate the charge exchange  forces between members of the isodoublet, the
neutron and proton  and  the neutrino and the electron.  In making this step
Klein discards the $\bar \Psi  \gamma^5 \partial_5 \Psi$   term in the kinetic
energy
that would give  generate a large mass term (of order the Planck mass)
 for the charged fermion fields.  He remarks that this is  consistent with the
symmetries.  Actually it is not
consistent with the  five dimensional general
covariance.  But he did not care about that.  He was only trying to preserve
the
four dimensional symmetries.  So he throws  away the  mass term of the charged
fermions coming from their kinetic energy in the
  fifth dimension in order  to keep them degenerate with their neutral
partners.

Thus Klein  has a wave equation for matter   coupled to charged gauge
fields plus  neutral  gauge fields. Thus,
 in addition to electromagnetism mediated by
the photon, he gets other forces that  identifies as the  nuclear forces
consisting of charge exchange between protons and neutrons and charge exchange
between protons
and neutrons and between electrons and neutrinos accounting for  both  the
strong and the weak nuclear forces.
\vskip -0.5 truein
\hskip -1.0truein {\epsffile{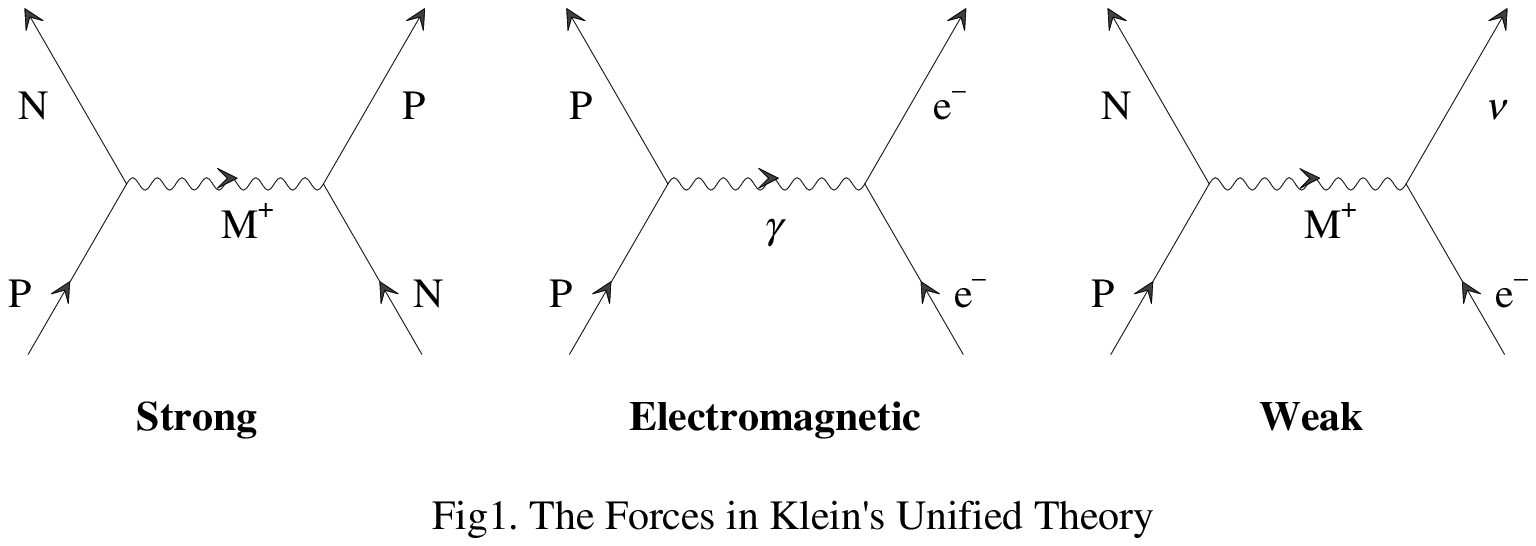}}

Next  Klein turns to the   field  equations for the gauge bosons?
The  method was well known to Klein.  He simply
took the Einstein    action in five dimensions, $\kappa\int d^5 x \sqrt{g} R_5
$,  and   reduced it to four dimensional form. As in the standard Kaluza-Klein
theory the Lagrangian reduces   to the four-dimensional Einstein Lagrangian
plus the square of the gauge field strength.  The  gauge field   is  given by
the
usual commutator of the covariant derivatives that  that appeared in the Dirac
equation
$ \nabla_\mu = \partial_\mu - \sqrt{\kappa} \chi_\mu \partial_5$.
This lead him to the Lagrangian for the gauge bosons, neutral and charged
\begin{eqnarray}
{\cal L_{\rm gauge} }& = &  -{  \textstyle {1\over 4} } \left( A_{\mu \nu}
A^{\mu \nu}
+  B_{\mu \nu} \bar B^{\mu \nu}  \right);\ B_{\mu \nu } = \left( \partial_\mu -
{ie \over \hbar c} A_\mu \right)B_\nu -\left( \partial_\nu - {ie \over \hbar c}
A_\nu \right)B_\mu
  \nonumber \\ A_{\mu \nu} &=& \partial_\mu A_\nu -\partial_\nu A_\mu +{ie
\over \hbar c}\left(
   B_{\mu } \bar B_{  \nu} - B_{\nu } \bar B_{  \mu}\right),
\end{eqnarray}
 from which
he derived the field equation for the gauge bosons.  The action for the
charged mesons $B_\mu$ involves the
  minimal coupling of these charged bosons to
electromagnetism.  A  new feature, which he points out,  was the typical
Yang-Mills  term $\left(
   B_{\mu } \bar B_{  \nu} - B_{\nu } \bar B_{  \mu}\right)$ in the
electromagnetic
field strength coming from the contribution of the charged vector bosons to
the electric current.    This looks a lot like Yang-Mills
theory.  But actually it is not.

At this point Klein adds   a mass term, $M^2 \bar B_\mu B^\mu $ as well, in
order that the  $B$ mesons can be identified with  the massive mesotrons of
Yukawa.   The mass term   he adds  is consistent, he
says, with gauge invariance.  It is of course not consistent with non-Abelian
gauge invariance but he was not thinking
about non-Abelian gauge invariance. The mass term {\em is}   consistent with
electromagnetic gauge invariance.

 Klein was
 not trying to construct an $SU(2)$ gauge theory.  He was just trying to
construct the $U(1)$ gauge theory of charged mesotrons, so the mass term was
allowed.  In fact Klein almost did construct   an $SU(2)$ gauge theory; but he
not quite.
The reason had to do with  hypercharge. If we write the $2\times 2$   vector
meson matrix in more conventional form, so that the coupling to the nucleon
iso-doublet   is
$\bar \Psi  \gamma_\mu W^\mu \Psi$, we see that
\begin{equation}
W_\mu= B^1_\mu \sigma_1 +B^2_\mu \sigma_2  + A_\mu {1-\sigma_3 \over  2}=
\left( \matrix{0 & B_\mu \cr \bar B_\mu & A_\mu \cr     }\right),
\end{equation}
These generators: $\sigma_1, \sigma_2$ and   ${1-\sigma_3 \over  2}$  are not
the generators of $SU(2)$.  Correspondingly the action he writes down
for the gauge bosons   is not the $SU(2)$ Yang-Mills action, there is a factor
of 2 wrong.
 This of course is well known to us today. To construct  gauge theory with one
iso-doublet requires a gauge group of $SU(2)\times U(1)$, as in the standard
model of Glashow, Weinberg and Salam. If you want a gauge theory with only one
neutral gauge
boson, the photon field, than one must put the matter into triplets of $SU(2)$,
as in the Georgi--Glashow model. Klein followed neither  of these approaches,
since he was not trying to construct a non-Abelian gauge theory. So he almost
invented  $SU(2)$ gauge theory but not exactly.   However,  he was very close.

Klein ends his discussion  by  making a few remarks about the quantization of
his   theory. He notes that one
can quantize this theory in the same way as  electromagnetism.  There is the
usual problem of a singular Lagrangian but he says that  Rosenfeld has solved
that
problem for
for QED and one  can do the same for his theory.  Actually, we know that it
is much more complicated  to quantize such theories, but he did not  know that.

Unlike all the other talks, the discussion  following Klein's talk was very
short. There was only one
remark by Moeller.  Moeller  noted  that there was  recent
experimental evidence for a neutral component
of the nuclear force.  The exchange of a neutral heavy Yukawa  meson   does not
seem to appear in your theory, Mr Klein, so what are you
going to do about that?    Klein answered that the cure is   simple enough; he
will just    add to the $2\times 2 $  gauge field a new diagonal component,
\begin{equation}\chi_\mu = \left(  \matrix{ A_\mu  & \bar B_\mu \cr B_\mu  &
A_\mu   \cr }\right) \to
  \left(  \matrix{ A_\mu -C_\mu& \bar B_\mu \cr B_\mu  & A_\mu + C_\mu \cr
}\right) .
\end{equation}
The extra  neutral  gauge field $C_\mu$, he says should have no $x^5$
dependence and  can be given  any mass you want. The exchange of this new
vector boson
might explain the neutral   nuclear force but, as he
honestly remarks,  being a vector particle it will be repulsive and not
attractive.   I regard this on-the-spot answer as quire remarkable.  is is sort
of a generalization of $SU(2)$ to $SU(2)\times U(1)$,
which, as you all know, was the step made 30 years later which gave rise to
the modern electroweak theory.

There was no further discussion.  Clearly the talk
 was over everyone's  head and might have been regarded, even in
comparison to Eddington's theory, as totally outlandish.
{}From our point of view, over fifty years later, it seems  remarkable how
reasonable
were the assumptions   that he made   and it seems amazing
how close he came to the truth.

\section{\bf Conclusions}

  Why did Klein's theory have
  no impact on the development of   physics?   There are many
possible reasons.    First, is that it is clear that Klein did not completely
understand
what he had done, a common phenomenon    among pioneers who often make great
leaps of imagination but do not appreciate the revolutionary aspects of their
creations (a good case is  Planck and the quantum theory.)    Klein's goal was
to
construct a theory of all the
forces based on a $U(1)$ gauge theory of iso-spinors.  He almost constructed
an  $SU(2)$ gauge theory, but not exactly.  I do not think he really
understood that he even came close to it; that was not his concern.   Second,
Klein never published a paper on this theory. As we have learned from
Professor Pais he wrote to Bohr  for his advice on publication.
There is no evidence of a response from Bohr and for some reason Klein
did not go ahead and publish.  So his new  ideas were buried   in the rather
obscure proceedings of the  Warsaw conference.
Finally,  the second world war broke out and
Klein was isolated from the community of physicists who  were was off doing
other things. By the time the war ended and people got back to doing this kind
of physics
he had probably forgotten what he had done.

Could it have been different?  Looked at from afar Klein's attempt at a unified
theory of the forces of nature in 1938 looks very similar to the successful
theory of elementary particles that was completed in 1973, a non-Abelian gauge
theory of the electro-weak and strong interactions, based on the gauge group
$SU(2)\times U(1)\times SU(3)$.    Could the route  from Klein's outline of a
gauge theory of nuclear forces to the standard model been more direct?  Is it
possible that if Klein had published his paper or
gone on the lecture circuit, people would have found these ideas  fascinating
and
started to really understand gauge theories and developed the standard model
earlier?
   Probably not.  It seems inconceivable that one could have arrived at the
standard model without going through the long succession of experiments of the
1950's and 1960's, accompanied by the many attempts at theoretical model
building. The actual path to the standard model was indirect and based on trial
and error. The experiments were crucial to this development. They revealed the
small deviations from Dirac's relativistic atom that stimulated the development
of quantum field theory and the understanding of renormalization; the existence
of a whole series of hadronic resonances that suggested the composite nature of
hadrons; the elucidation of the symmetries, good and bad, of the weak
interactions and the V-A nature of weak currents; the discovery of Yang-Mills
theory; the approximate $SU(3) \times SU(3)$ symmetry of the strong
interactions which; led to the hypothesis of quarks and color; the
understanding of chiral symmetry  and spontaneous symmetry
breaking and the Higgs mechanism that led to the electro-weak theory; the
discovery of scaling in deep-inelastic scattering which led to the discovery of
asymptotic freedom and the proposal of QCD. As should have been clear from my
summary of
the rest of the conference, the knowledge of particle physics  in 1938 was
incredibly primitive and the knowledge of quantum field theory was
equally primitive. The  experiments that were being carried out in 1938 were at
energies of  a few   MeV
at best.  It was simply premature to attempt to develop a theory of the nuclear
force when the   characteristic scale of the strong interactions is a
100 MEV to 1GeV and the characteristic scale of the weak interactions is a
100 GeV.   One required detailed experimental exploration at energies
well above the characteristic mass scale of the the relevant interactions
before things became clear.

What is  the lesson of all of this for us now?   Today we have a theory of  all
the forces of nature that we observe,  just as Klein wanted,  that agrees with
all experiments up to energies of a TeV or so    with impressive accuracy.
Many theorists are trying, as Klein did to extrapolate to territory unexplored
by experiment. Do we have any more chance of  succeeding than did Klein?

  Theorists are pretty good at extrapolating from what they already know to
guess where new physics might arise, where problems will appear, where new
thresholds will show up;
even if they are bad at guessing the new physics at these thresholds. Thus,
after the Fermi theory of the weak interactions one could easily guess that
there had to be new physics at $100$ GeV, even though one had no good idea as
to what the new physics would be. As far as we can tell, if we use the standard
model to extrapolate the known forces, we find that new physics---fundamentally
new thresholds---will only appear at extraordinarily high energies, 17 orders
of magnitude removed from present  day experiment.

Of course there are likely to be many new experimental discoveries in between
the TeV region and the Planck energy. All unified theories, certainly string
theory, predict that there will be much new stuff in this region. But  the
truly new phenomenon  that might indicate a fundamental modification of the
laws of physics might not be seen until the unification or Planck scale.

   Can we  succeed in making this extrapolation.  One can easily give arguments
both pro and con.   The arguments against success are easy---history teaches us
that without direct experimental clues and tests theorists tend to go wrong.
Klein's example is a good
case of how one can be so close to the truth, yet so far from true
understanding.
However there are some differences between the situation today and that in
1938. One is that we have, unlike Klein, an extremely solid
spring board.  Klein did not have a theory that explained everything that
was observed at his time from which he was trying to extrapolate to higher
energy. It is not easy to extend such a theory without contradiction, so
consistency is  a guide.     Also, the extrapolation, when
  measured  not in terms of energies   but in terms of inverse
couplings (theoretically the correct way to measure energies) is not
such a big extrapolation.  On an inverse coupling scale,  going from  1 MeV
physics to
energies of order  the $W$ mass is the same as going from the $W$ mass to the
Planck
scale.  This is a big extrapolation but not unprecedented.
And finally we have  the incredible luck of knowing a bit of Planck scale
physics---namely  gravity.
 We are therefore presented with the
obvious challenge to understand that  part  of Planck scale  mass physics,
together with trying to   unify   the electro-weak and strong interactions.

In any case, in my opinion, we have no choice but  to try. We must  emulate
Klein and
be daring.
 
\end{document}